# Scanning Tunneling Spectroscopy of Graphene on Graphite

Guohong Li, Adina Luican & Eva Y. Andrei

*Department of Physics and Astronomy, Rutgers University, Piscataway, New Jersey 08854*

Abstract

We report low temperature high magnetic field scanning tunneling microscopy and spectroscopy of graphene flakes on graphite that exhibit the structural and electronic properties of graphene decoupled from the substrate. Pronounced peaks in the tunneling spectra develop with field revealing a Landau level sequence that provides a direct way to identify graphene and to determine the degree of its coupling to the substrate. The Fermi velocity and quasiparticle lifetime, obtained from the positions and width of the peaks, provide access to the electron-phonon and electron-electron interactions.

Graphene, a one-atom thick form of crystalline-carbon [1-3] possesses extraordinary electronic properties owing to charge-carriers that move like ultra-relativistic particles (massless Dirac-fermions). Their relativistic nature becomes evident in the presence of a magnetic field through the appearance of an unusual Landau level (LL) sequence [2] with square-root dependence on field, B, and level index *n* together with a unique field independent *n=0* level at the Dirac point (DP). One of the first manifestations of the *n=0* level was the appearance of half integer quantum Hall plateaus [4,5]. Subsequently, scanning tunneling microscopy (STM) of graphene deposited on insulating substrates [6-9] and zero field scanning tunneling spectroscopy (STS) showed surprising deviations from theoretical expectations prompting concerns over the interpretation of the STS due to the poor screening provided by the insulating substrate [10].

Here we study graphene flakes deposited on a graphite substrate which provides the necessary screening to ensure reliable STM and STS. We demonstrate that in this system it is possible to access the intrinsic properties of the quasiparticle carriers in graphene and to determine their degree of coupling to the substrate. In zero field we find that the density of states (DOS) is linear and vanishes at the DP as expected of massless DF [2]. In finite field we observe the appearance of a sharp sequence of Landau levels (LL) with square root dependence on field and LL index together with the distinct *n=0* level. We further show that the DFs are slowed down by electron-phonon interactions and their lifetime is limited by electron-electron interactions.

Experiments were conducted in a home built, low temperature (T = 4.4K for this work) high field STM [11] using mechanically cut Pt-Ir wire tips. The tunneling conductance, dI/dV, was measured by lock-in detection with 340 Hz bias voltage modulation. Samples were prepared by cleaving highly-oriented pyrolitic-graphite (HOPG) and the coarse motors of the STM were used to spot graphene flakes.

Once a flake is found its structure and electronic properties are studied to determine its degree of coupling to the graphite substrate. STM topography of such a flake, Fig. 1(a), reveals three layers separated in height by atomic steps. The top layer in region A exhibits the distinctive honeycomb structure, while that in region B is triangular as expected of Bernal stacked graphite [12]. Thus the images in Fig.1 suggest that in B the top layer is coupled to the underlying layers, while in A, separated from B by a macroscopic ridge-like defect, the top layer is a single layer of graphene (SLG) decoupled from the substrate. Further confirmation that region A consists of SLG is gained by accessing the DOS through STS [13].

Due to the unique band structure of graphene the DOS in SLG is linear in energy and vanishes at the DP[2]. This is not the case when two or more graphene layers are coupled together because interlayer coupling produces additional states at the DP leading to a finite DOS there. Therefore a linear DOS that vanishes at the DP is a signature of massless DF and can thus be used to identify SLG decoupled from the substrate. A more stringent test of isolated SLG is to apply a magnetic field normal to the layer and probe for the appearance of a sequence of LL that is unique to massless DF [2]:

$$E_n = E_D + \text{sgn}(n)\sqrt{2e\hbar v_F^2 |n| B}, \qquad n = ...-2,-1,0,1,2,3... \qquad (1)$$

where $E_D$ is the energy at DP, $v_F$ the Fermi velocity, $-e$ the electron charge, with $n>0$ corresponding to electrons and $n<0$ to holes. Any coupling to the substrate introduces extra levels and eventually smears out the LL sequence [14,15,29] Thus the square-root dependence of the LL energy on both $B$ and $n$ uniquely identifies decoupled graphene layers and was used in our search of decoupled flakes. Several completely decoupled flakes were found. Most of the data reported here is on one sample that was studied most extensively.

In Fig.2(a) we show the bias dependence of the tunneling conductance, a quantity proportional to the local DOS [13]. In A the DOS is V-shaped and vanishes at the DP, as expected for SLG and in agreement with the honeycomb structure observed in topography. We note that the DP is offset by ~ 16meV above the Fermi energy (identified with zero bias) indicating unintentional doping with a concentration of $\sim 2 \times 10^{10} cm^{-2}$ hole carriers. In contrast the DOS in B remains finite at all energies indicating coupling to the substrate. The difference between the two regions becomes evident in the low energy tunneling conductance map, Fig.2(b), showing that the ridge acts as a border between a region of decoupled SLG and one of coupled layers.

To further ascertain that region A corresponds to decoupled SLG we carried out STS in the presence of a magnetic field normal to the surface. As shown in Fig.2(c) the spectra develop into a sequence of well-defined LL peaks with increasing field. In A the peaks are substantially more pronounced than the peaks reported in previous studies [11,15] where the spectral weight of LL was reduced by substrate coupling. The massless Dirac fermion character of the spectra in A is revealed by plotting the peak energies against the reduced variable $(|n|/B)^{1/2}$. This scaling procedure collapses all the data unto a straight line, Fig.2(d), attesting to the Dirac fermion nature of the charge carriers. Comparing to Eq.(1) we obtain (from the slope) the value of the Fermi velocity, $v_F$ = 0.79 ×10$^6$ m/s. In addition we find the position of the Dirac energy relative to the Fermi level, $E_D$ = 16.6 meV is consistent with the zero-field data, Fig.2(a). The scaling of the LL sequence holds across the entire sample in A [16]. In contrast, no such collapse is seen in B where graphene is coupled to the substrate.

The value of $v_F$ obtained from the level sequence is ~20% below the tight-binding value [2] and than that measured in weakly coupled layers on graphite [11,19]. Below we show that this result can be attributed to electron-phonon (e-ph) interactions [21,22]. According to density functional theory (DFT) calculations [24], e-ph interactions introduce additional features in the electron self-energy, leading to a renormalized velocity $v_F = v_{F0}(1+\lambda)^{-1}$ and to dips in the renormalization factor, $(v_F - v_{F0})/v_F$, at energies $E_F \pm \hbar\omega_{ph}$. Here $v_{F0}$ is the bare Fermi velocity, $\lambda$ the e-ph coupling and $\hbar\omega_{ph}$ the energy to the most strongly coupled phonons. The e-ph coupling also introduces a kink in the zero field DOS at $\hbar\omega_{ph}$. Examining the zero field spectra in Fig.2(a), we note a strong shoulder-like feature around ±150 mV in the spectra in region A (but not B), consistent with the e-ph coupling scenario. A more detailed analysis of the e-ph coupling is discussed next.

The LL sequences provide strong evidence that the quasiparticle excitations in graphene are 2d massles Dirac fermions together with the average Fermi velocity for energies up to 150 meV. Using these facts and the linearity of the DOS [2], we obtain the zero field calibration of the DOS, $\rho(E) = \dfrac{3^{1.5}a^2}{\pi}\dfrac{|E-E_D|}{\hbar^2 v_F^2} = 0.123|E-E_D|$, for energies below 150 meV ($a$ = 1.42 Å).

Comparison between this DOS and the normalized zero field tunneling spectra in Fig.3(a) yields a remarkable consistency justifying the assumption dI/dV $\propto$ DOS [13] within the range −150 mV to 150 mV. Assuming this proportionality remains valid to higher energies (~ 600 meV) and an isotropic band (valid in this energy range where trigonal warping is negligible) we obtain:

$k(E) = \pm\left|\dfrac{\pi}{A_c}\int_{E_D}^{E}\rho(E')dE'\right|^{1/2}$ by integrating the measured differential resistance curve, and from it the dispersion relation $E(k)$ shown in Fig.3(b). Here $A_c$ is the area of a unit cell. This allows to calculate the energy dependence of the Fermi velocity, $v_F \equiv \dfrac{dE}{\hbar dk}$, shown in Fig. 3(c). Identifying the asymptotic value of the velocity with $v_{F0}$ we obtain $\lambda \sim 0.26$. The curve bears a striking similarity to that obtained by DFT [24]: it exhibits two dips at the energy of the $A_{1g}$ phonon $\pm 150(\pm 20)$meV, suggesting that this phonon, which couples the $K$ and $K'$ valleys (Fig.3(b) inset) and undergoes a Kohn anomaly, is an important player in the velocity renormalization. Further support for the role of e-ph coupling comes from ARPES spectroscopy of graphene on SiC which revealed similar kinks in the dispersion and a down renormalization of $v_F$ [25].

We now turn to high resolution (2mV) LL spectroscopy at 4 T shown in Fig. 4(a). The sharpness and high definition of the LL peaks made it possible to extract the energy dependence of the quasiparticle lifetime. The level sequence can be fit with a sum of peak functions centered at the measured peak energies with the line width of each peak and an overall amplitude as free parameters. Interestingly, we found that Lorentzian fits were significantly better than Gaussians suggesting an intrinsic lifetime rather than impurity broadening. We focus on the hole sector in Fig.4(b) where the energy range covered by the LL peaks is wider than on the electron side. The line widths were found to increase linearly with energy corresponding to an inverse lifetime:

$$\frac{1}{\tau_n} = \frac{|E_n|}{\gamma} + \frac{1}{\tau_0},$$ where $E_n$ is the LL energy in units of eV, $\gamma \sim 9$ fs/eV and $\tau_0 \sim 0.5$ ps at Fermi level. The latter translates into a mean free path of $l = v_F \tau_0 \sim 400$ nm which is comparable to the sample size.

Several mechanisms leading to a finite quasi-particle lifetime are discussed in the literature. 1) Disorder induces a finite elastic lifetime at low energies and leads to decreasing LL width away from the Fermi level [27], a trend which is opposite to what is observed here. This is also consistent with our STM topography that shows no signature of structure disorder in the area where the tunneling spectra were taken. 2) Broadening due to electron-phonon scattering is expected to be negligible for energies E < 150meV, below the typical $A_{1g}$ phonon energy, the main player in the velocity renormalization [24]. 3) Electron-electron interactions in graphene give rise to a linear energy dependence of the inverse lifetime characteristic of non-Fermi liquid behavior [28]. Theoretical estimates for the electron-electron interaction limited lifetimes in zero field give $\gamma \sim 20$ fs/eV. Since the electron-electron interactions are enhanced in magnetic field, it is possible that the agreement would be even better with calculations made in finite field.

A new feature revealed in the high resolution spectrum, Fig.4(a), is the splitting of the n=0 LL. As we show in Fig.4(d) the splitting is of order ~ 10 mV and depends rather weakly on field. High resolution zero field spectroscopy, Fig.4(c), reveals a 10 mV gap at the DP suggesting that it may be related to the splitting in finite field. The fact that the gap opens at the DP rather than the Fermi energy implies that it is driven by broken sublattice symmetry, but more work is needed to elucidate the symmetry breaking mechanism and the splitting of the n=0 level in magnetic field. Work is in progress to address these questions.

The tunneling microscopy and spectroscopy experiments presented here demonstrate that graphene flakes deposited on graphite can be sufficiently decoupled from the substrate so as to exhibit the structural and electronic properties expected of pristine graphene: a coherent honeycomb structure extending over the entire sample and a V shaped density of states that vanishes at the Dirac point. In a magnetic field the development of a single sequence of pronounced Landau level peaks corresponding to massles Dirac fermions provides compelling evidence of the decoupling. The LL spectra give access to important many-body effects including electron-phonon interactions which produce a downward normalization of the Fermi velocity, and electron-electron interactions reflected in quasiparticle lifetimes that are inversely proportional to energy.

We thank E. Abrahams, A. V. Balatsky, A. H. Castro Neto, J.C. Davis and D. Langreth for useful discussions. Work supported by DE-FG02-99ER45742, NSF-DMR-045673 and Lucent.

Figure Captions

FIG. 1 (color online). Topography of graphene layer isolated by extended defects on a graphite surface. (a) Large area topography. Two underlying defects are seen: a long ridge that runs diagonally under the top two layers and a fainter one under the first layer (dashed line). The long ridge separates a region with honeycomb structure (region A) above it, Fig.1(e), from one with triangular structure (region B) below, Fig.1(f). Two arrows mark positions where atomic resolution images were taken. (b) High resolution image where the fainter ridge is visible. (c) Cross-sectional cut along line $\alpha\alpha$ in (a) showing that the separation between the top and second layer is larger than the equilibrium value (0.34 nm) near the fainter ridge. (d) Cross-sectional cut along line $\beta\beta$ showing that the height of an atomic step far from the ridges is comparable to that in Bernal-stacked graphite. (e,f) Atomic-resolution image showing honeycomb structure in region A (atoms visible at all 6 hexagon vertices) and triangular structure in B (atoms seen only at 3 vertices corresponding to only one visible sublattice). A coherent honeycomb structure is is seen over the entire region A. Set sample bias voltage and tunneling current were 300mV and 9pA for (a), 300mV and 49pA for (b), 200mV and 22pA for (f), 300mV and 55pA for (e), respectively.

FIG. 2 (color online). Spectroscopic evidence for de-coupled graphene layer. (a) Zero field spectroscopy taken in regions A and B marked by squares in Fig.2(b). Top panel shows that the tunneling conductance for decoupled graphene (region A) is V-shaped and vanishes at the DP (marked by the arrow). A typical spectrum for graphite is included for comparison. Bottom panel shows that in the more strongly coupled layer (region B) the differential conductance does not vanish at the DP. (b) Differential conductance (dI/dV) map at the DP energy as marked by arrows in Fig.2(a). The scanned area is the same as in Fig.1(b). dI/dV vanishes in the dark region but is finite in the bright region. (c) Field dependence of tunneling spectra in region A showing a single sequence of Landau levels.(LL) The peaks are labeled with LL index n. Each spectrum is shifted by 80 pA/V for clarity. (d) LL energy showing square-root dependence on level index and field. The symbols correspond to the peaks in Fig.2(c), and the solid line is a fit to Eq. (1). From the slope we obtain $v_F = 0.79 \times 10^6$ m/s and from the intercept $E_D$=16.6 mV indicating hole doping. The tunneling junctions parameters were set at a bias voltage of 300mV and tunneling current of 20pA for (a) and (c) and 53pA for (b). The amplitude of the ac bias voltage modulation was 5mV for all spectra in the figure.

FIG. 3 (color online). Zero field tunneling spectra, energy-momentum dispersion and Fermi velocity. (a) Tunnelling spectra and density of states (see text). Thick line is the DOS calculated using $v_F$ obtained in Fig.2(d). Note the consistency of spectra for tunnelling resistance varying from 3.8 GΩ to 50 GΩ. Legend shows tunnelling junction settings. (b) Energy-momentum dispersion obtained as described in the text. Inset: Diagram of intervalley scattering mediated by $A_{1g}$ phonon. (c) Energy dependence of Fermi velocity.

FIG. 4 (color online). LL sequence, quasi-particle lifetime and gap at the DP. (a) High resolution tunneling spectrum at 4T showing the LL. The dashed line marks the DP obtained by fitting the non-zero LL sequence. The n=0 LL splits into two peaks, 0- and 0+. (b) Zoom into LL sequence in hole sector. Solid squares are data points and the line represents a fit [26] with Lorentzian peaks. Inset, energy dependence of the peak width. (c) High resolution zoom into zero field spectrum showing the appearance of a ~10meV gap at the DP and an anomaly at the Fermi energy. (d) (top) Field dependence of the 0- and 0+ Landau levels. Solid symbols: data from Fig.4(a); open symbols from Fig.2(c). (bottom) Field dependence of splitting. Experimental settings are shown in the upper-left corners of Figs.4(a) and 4(c).

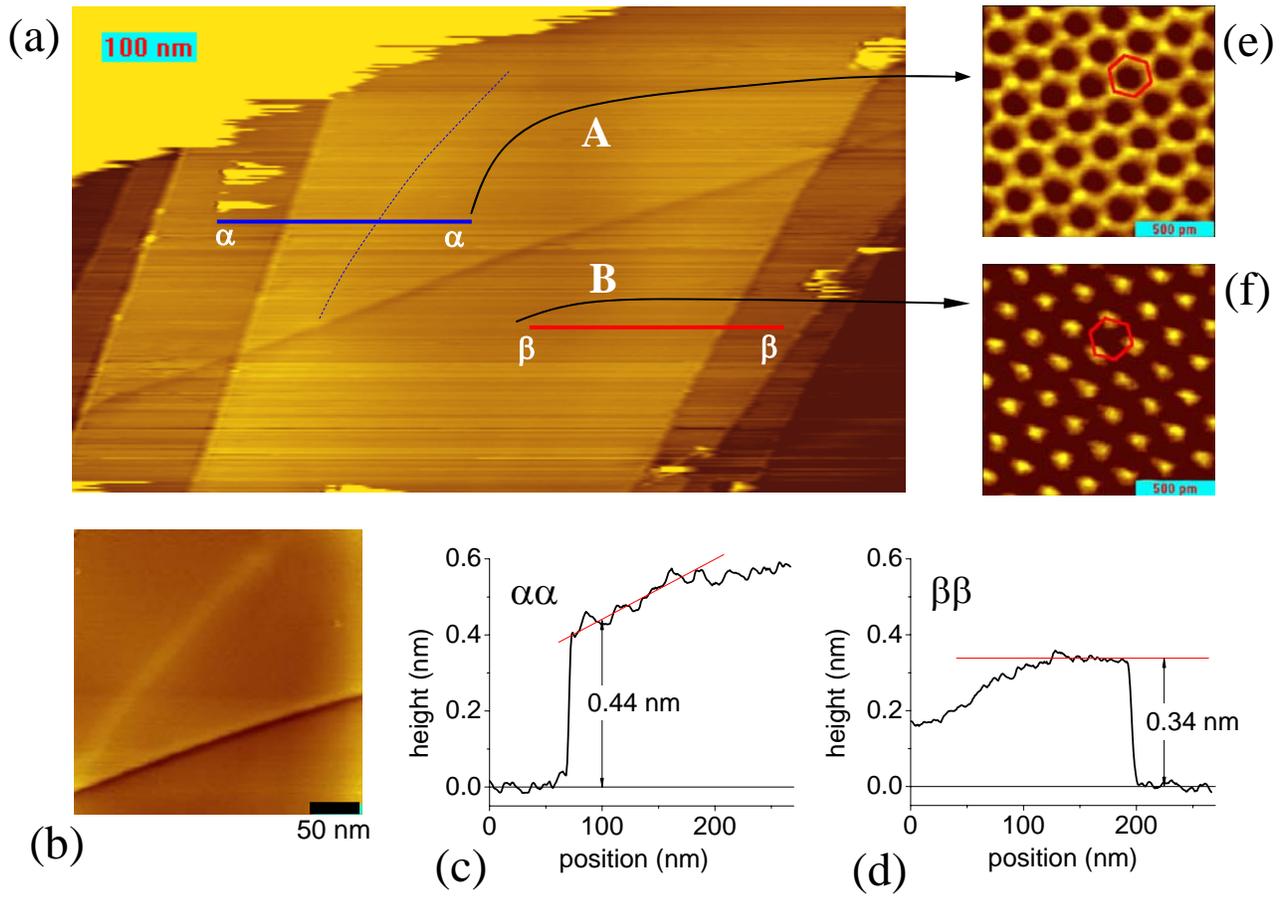

FIG. 1

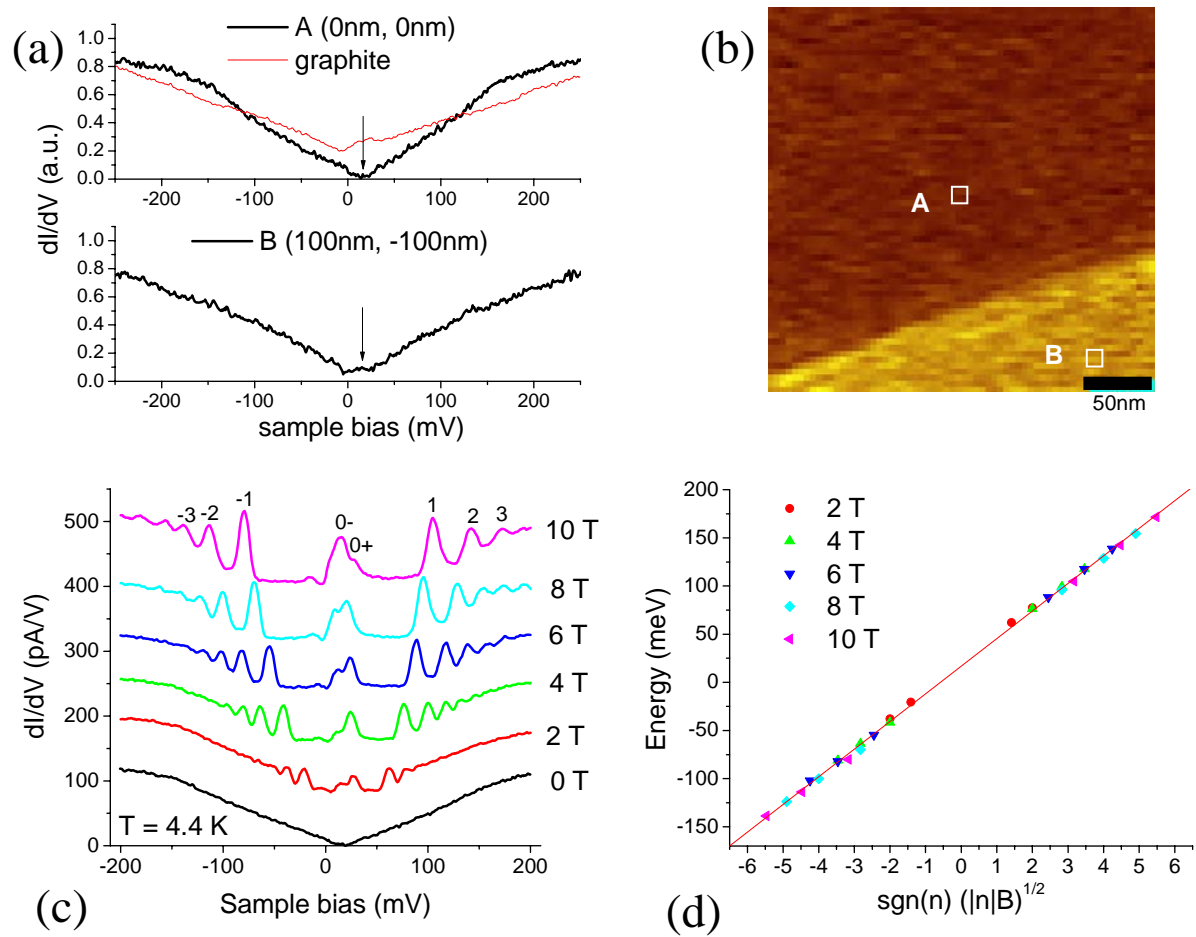

FIG. 2

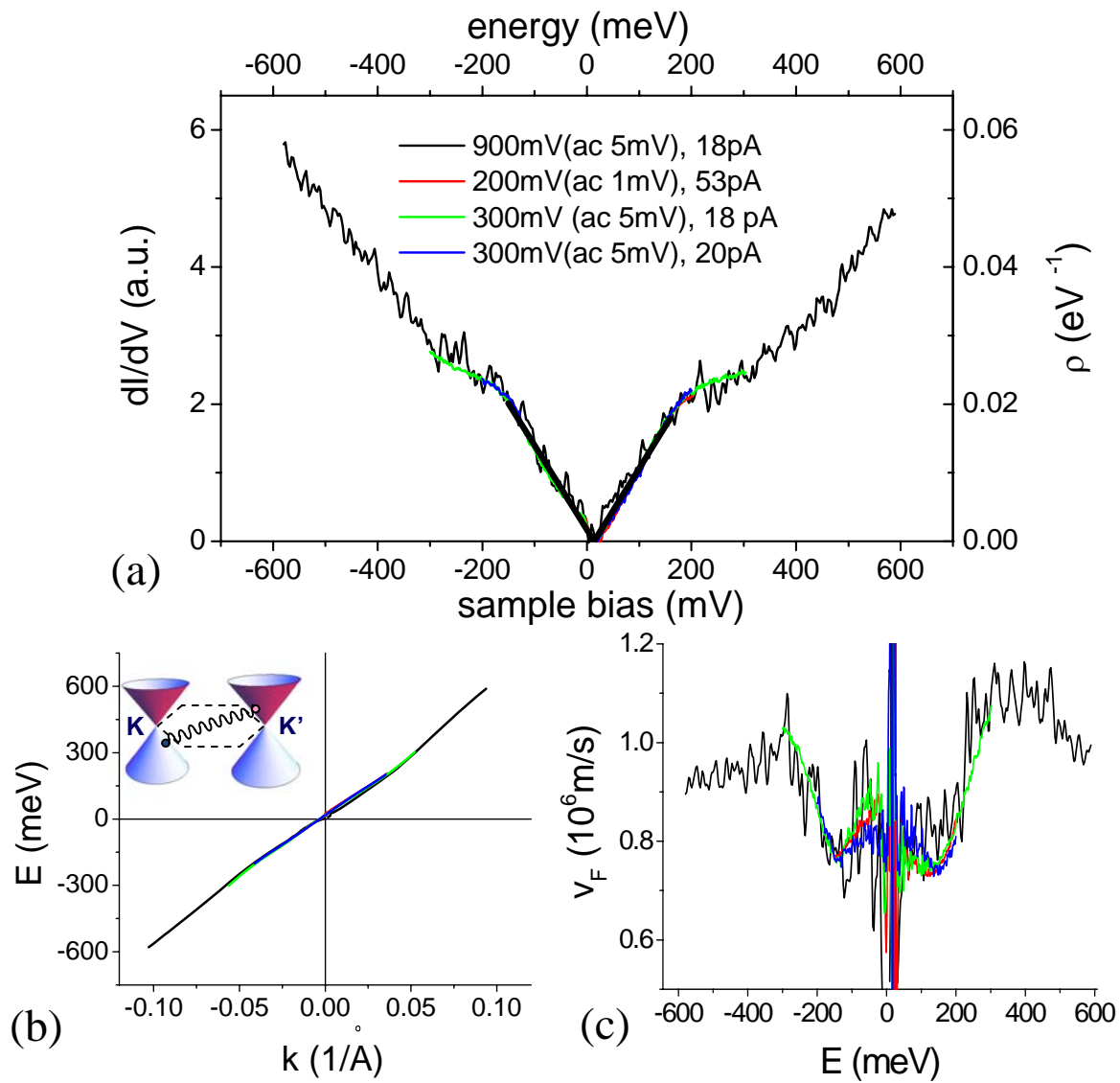

FIG. 3

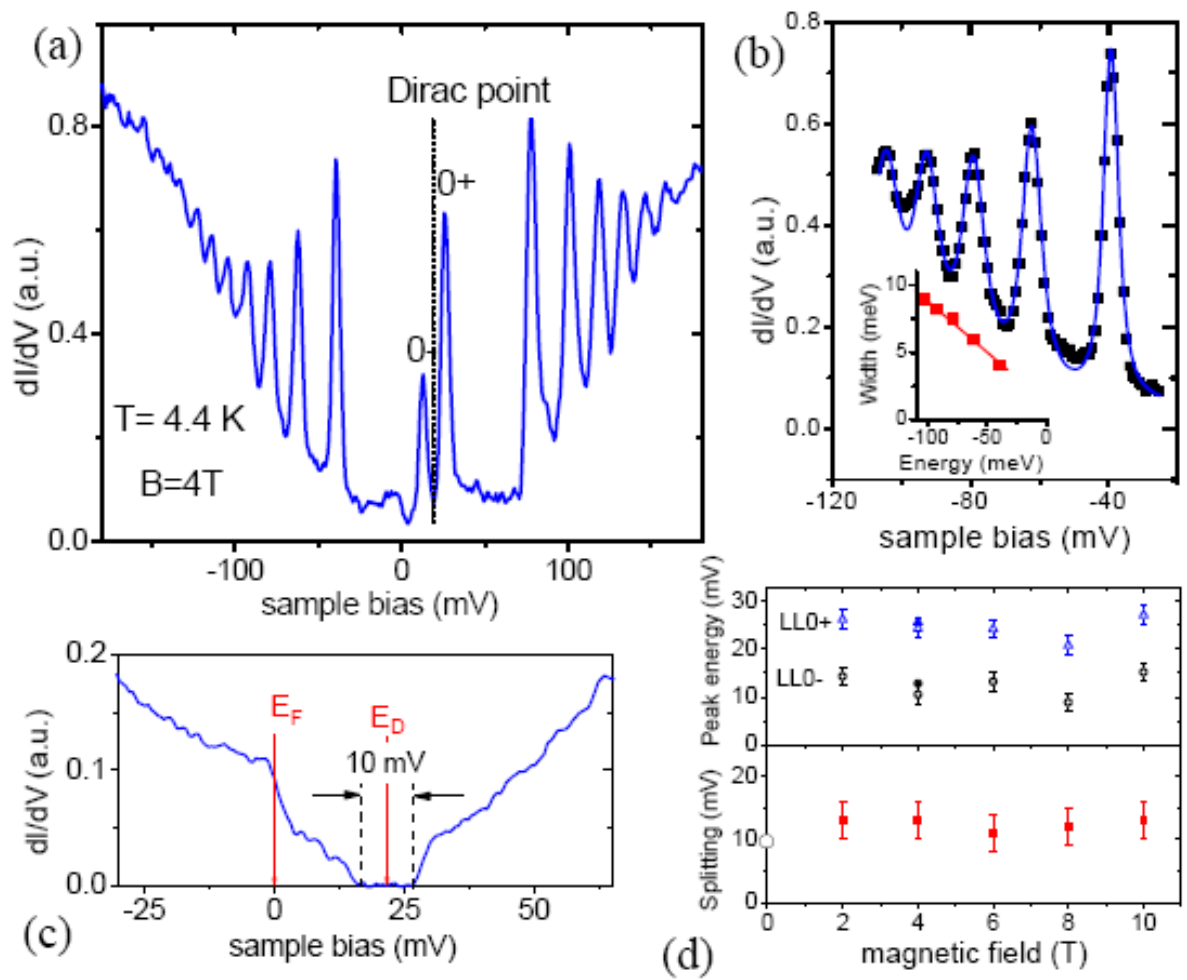

FIG. 4

Supplementary material for

# Scanning Tunneling Spectroscopy of Graphene on Graphite

Guohong Li, Adina Luican & Eva Y. Andrei

*Department of Physics and Astronomy, Rutgers University, Piscataway, New Jersey 08854*

1. **Searching for graphene on graphite surface**

    The findings presented in this work are the result of a systematic study on the surface of freshly cleaved HOPG in search of weakly coupled and decoupled layers of graphene. One of the main findings is that the appearance of well defined Landau level peaks provides a direct and unequivocal way to identify graphene and to determine the degree of its coupling to the substrate. In particular we found that on portions of the surface that are not decoupled representing high quality bulk graphite, Landau level peaks ( beyond the 0,-1 peaks), are barely observable consistent with previous reports on Kish graphite. On SPI-2 HOPG samples, every cleavage produced surfaces with a variety of defects including flakes, ribbons and steps. On these surfaces we developed an efficient procedure to find and identify weakly coupled graphene layers. Each run starts with a new Pt-Ir tip and a freshly cleaved surface that is subsequently studied extensively. Once the sample is prepared it is cooled to base temperature. Subsequently the surface is scanned in search of graphene flakes. We found that flakes that lie on top of extended defects are the most likely to be decoupled. These flakes are then characterized with topography followed by finite field spectroscopy in search of a well defined and pronounced sequence of Landau levels. We have found that the strength of the peaks is a good first indication of the degree of coupling: the weaker the coupling of a flake to the substrate the stronger the peaks. Typically a weakly coupled layer could be found by sampling an area of about 1μm a few times. In most cases more than one sequence of Landau levels is seen indicating a flake consisting of three or more layers that is decoupled from the bulk. A detailed description of the evolution of the Landau level sequence with coupling was reported in reference S1 and S2. In the work described here we focus on layers with extremely low coupling. For region B in figure 1 we were able to estimate that the coupling strength to the substrate was only about 40 meV, one order of

magnitude below the interlayer coupling of a perfect graphite crystal (S1,S2). Region A, which is the focus of the present manuscript, represents a completely decoupled graphene layer. In this region the peaks are strong and well pronounced, they constitute a single sequence of Landau levels whose energy exhibits a square root dependence on field and level index.

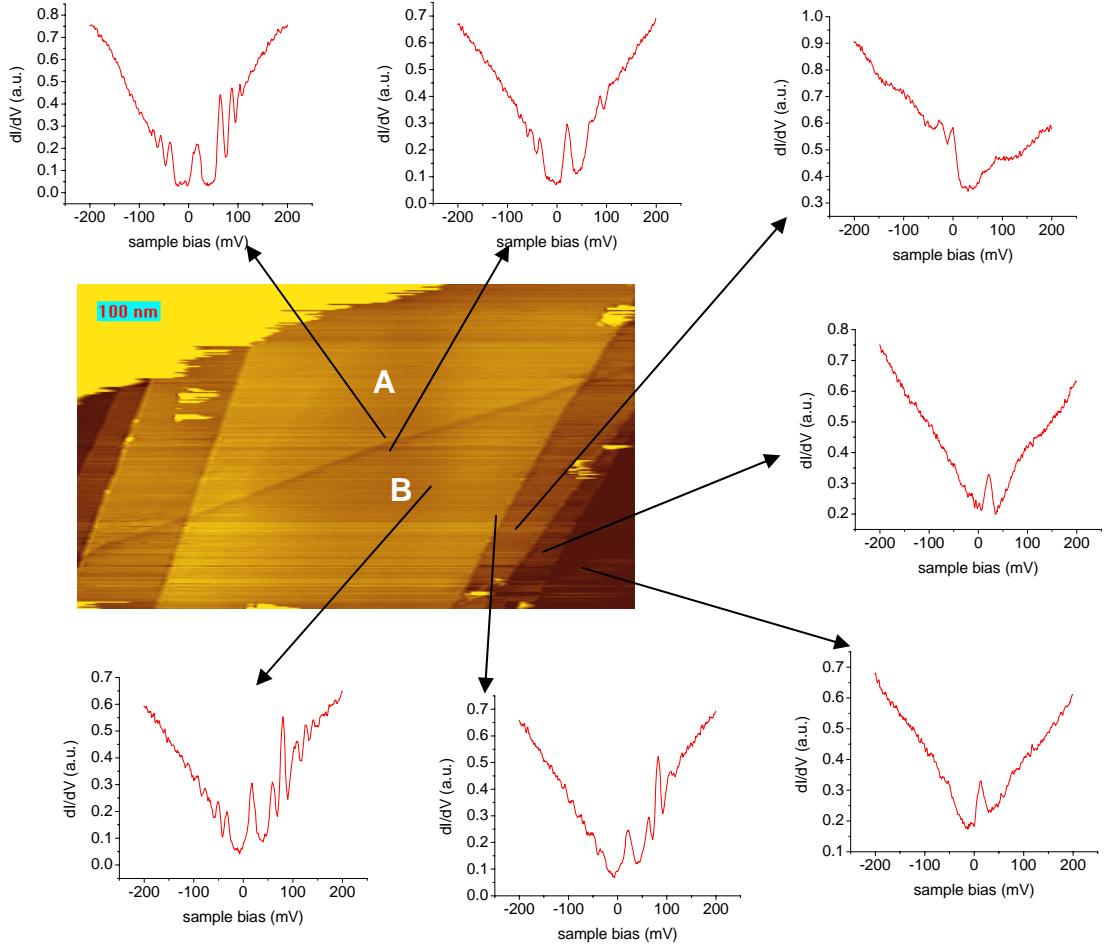

Fig.S1. Spatial dependence of tunneling spectroscopy on a graphite surface. The appearance of pronounced Landau levels indicates reduced coupling between the top graphene layers and the bulk. T=4.4K, B=3T, ac modulation 5mV in amplitude. Tunneling current was 20pA at sample bias voltage of 300mV.

In Fig.S1 we show the evolution of the Landau levels in a field of 3T from a region with strong coupling (lower right side of the figure) to region A where the graphene layer was decoupled. The spectra reported in the main text were taken near the center of the decoupled graphene to highlight the least disturbed electronic properties of graphene but the same Landau level structure was seen over the entire A region. We found that the Landau levels broadened over areas with disorder on the surface or near boundaries. The experiments were repeated by ramping the field up and down a few times and the systematic variation of the tunneling spectra with position and coupling were always reproduced. The experiments were typically run continuously over about 3 weeks and were usually discontinued following an accidental tip crash. During a run we had to retract the tip several times for transferring liquid Helium, and our STM was stable enough to find the same sample area with a drift less than 10 nm after transferring. Even after a tip-crash, we were able to treat the tip and qualitatively reproduce all the results.

## 2. Landau level width and quasiparticle lifetime

The LL sequence was analyzed using a sum of peak functions centered at the measured peak energies with the line widths and overall amplitude left as free parameters. The fits were carried out with both Gaussian and Lorentzian peak functions for the first 5 peaks so as to stay clear of the K-A' phonon energy. We find that the Lorentzian fits are consistently and significantly better than Gaussians as measured by comparing the standard deviations and as is obvious from a visual comparison of the fits shown in the figures below. Separate fits were done for the hole and electron sectors, with qualitatively similar results.

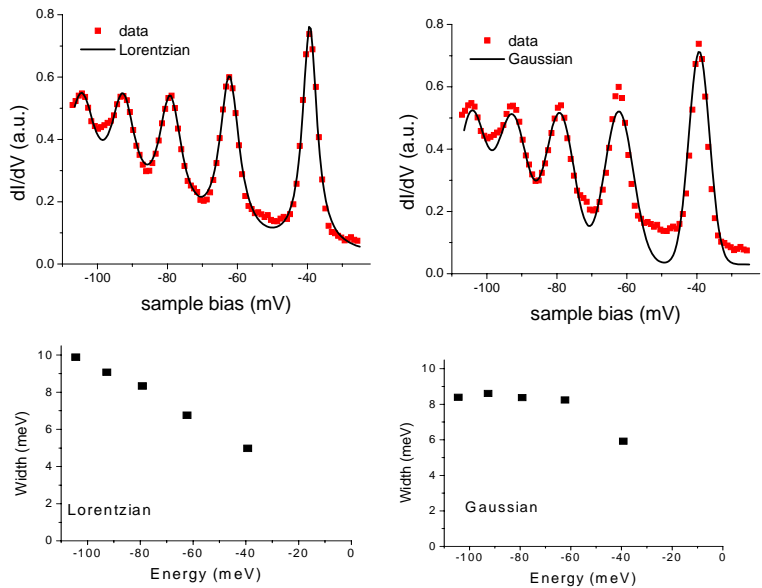

Fig. S2 Top- Landau levels at 4T fitted with Lorentzian (left panel) and Gaussian (right panel) peak functions. Bottom- Energy dependence of linewidth obtained from Lorenztian (left panel) and Gaussian (right panel) fitting.

*Effect of modulation amplitude*

In order for the measured linewidth to correctly reflect the intrinsic linewidth the ac modulation amplitude in the lock-in measurement should be small compared to the intrinsic linewidth. Below we calculate the effect of finite modulation.

If $f(x)$ is the intrinsic line shape to be measured and $h$ the amplitude of modulation, then the lock-in output is given by [26]

$$\sum_{n=0}^{\infty} \frac{h^{2n+1}}{2^{2n} n!(n+1)!} \frac{d^{2n} f}{dx^n}$$

which leads to the following substitution in fitting

$$f(x) \to f(x) + \sum_{n=1}^{\infty} \frac{h^{2n}}{2^{2n} n!(n+1)!} \frac{d^{2n} f}{dx^n} = f(x) + \frac{1}{8} f'' h^2 + \frac{1}{192} f'''' h^4 + ...$$

where for the fits discussed here $f(x)$ would be the Lorenzian or Gaussian peak functions without correction. The corrected fits with n up to 15 are plotted below. While the Lorentzian fit is somewhat improved using the reduced line width obtained from the new fits, the Gaussian peaks still cannot capture the line shape. The corrected Lorentzian fit and the deduced line width (also plotted in figure 4 of the main text) reveal that the linewidth is linear in energy within the measurement range.

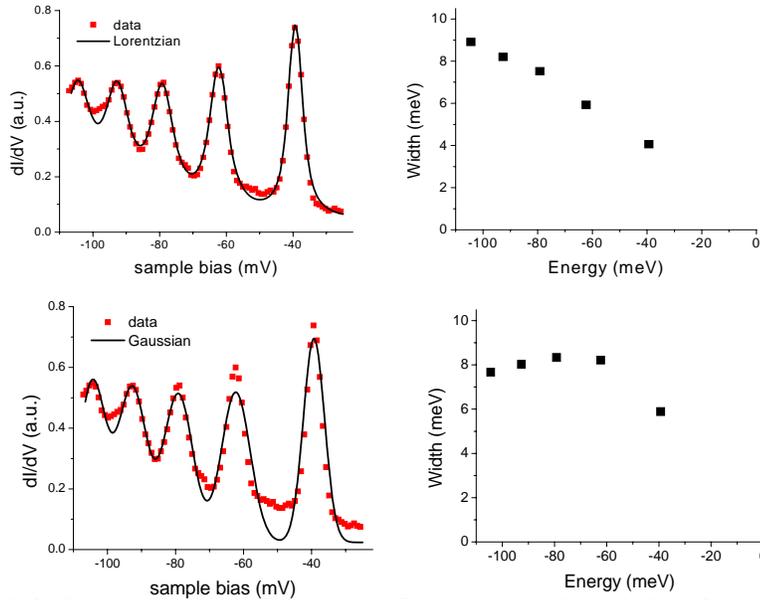

Fig. S3 Top- (left) Landau levels at 4T fitted with Lorentzian peaks including finite modulation correction. (right) Linewidth obtained from the fitting. Bottom-(left) Landau levels at 4T fitted with Gaussian peaks including finite modulation correction. (right) Linewidth obtained from the fitting.